# Oxygen incorporation in YBa$_2$Cu$_3$O$_{7-\delta}$ thin films: surface activation and degradation

Alexander Stangl*[1,2], Xavier Obradors[2], Anna Palau[2], Arnaud Badel[1,3], Teresa Puig*[2]

* alexander.stangl@neel.cnrs.fr, teresa.puig@icmab.es
[1] Université Grenoble Alpes, CNRS, Grenoble INP, Institut Néel, 38000 Grenoble, France
[2] Institut de Ciència de Materials de Barcelona (ICMAB-CSIC), 08193 Bellaterra, Barcelona, Spain
[3] Université Grenoble Alpes, CNRS, Grenoble INP, G2ELab – Institut Néel, 38000 Grenoble, France

Abstract:
The oxygen off-stoichiometry plays a pivotal role for the physical properties of superconducting oxides. Yet, there is a lack of knowledge on the fundamental processes of oxygen incorporation and correlated phenomena during oxygen post annealing treatments. Here, we deployed electrical probes ($\rho(T)$ and electrical conductivity relaxation (ECR) measurements) to gain a better understanding of the kinetics of the oxygen reduction reaction (ORR) in epitaxial YBa$_2$Cu$_3$O$_{7-\delta}$ (YBCO) thin films. We identified a new indicator for the onset temperature of oxygen incorporation and report for the first time drastic kinetic deactivation in bare YBCO. We demonstrate that surface decoration using silver micro islands both catalytically activates oxygen incorporation and bypasses surface degradation processes. Our results suggest that the ORR in the studied YBCO samples is limited by a surface reaction. Additionally, weak XRD signatures of the formation of extended bulk defects were identified, which have to be considered in the designing of an optimal oxygenation treatment to obtain best performing superconducting YBCO.

Introduction:

Oxide materials are considered game changers in the pursuit for a sustainable energy transition, due to their promising potential in the domains of energy harvesting, conversion, transport and storage, including oxide photovoltaics,[1,2] solid oxide fuel and electrolysis cells (SOFC & SOEC),[3,4] fusion-enabling high temperature superconductors (HTS),[5–8] and novel oxygen ion batteries.[9,10] For many members of this functional oxides materials class, the flexible oxygen off-stoichiometry, $\delta$, is vital to their physical properties.[11–13] This is especially true for superconducting oxides, such as YBa$_2$Cu$_3$O$_{7-\delta}$ (YBCO), where $\delta$ does not only govern the structural transition to the superconducting orthorhombic phase,[14] but also determines the superconducting critical temperature, $T_c$,[15] and strongly impacts the maximum critical current density, $J_c$,[16–20] via the release of electronic charges (charge doping) to the crystal upon oxidation. Thus, precise control of the oxygen stoichiometry is a pivotal step in the manufacturing of high performing superconducting materials, including so called coated conductors,[21] which represent the commercial solution to the strict crystallographic requirements to unfold the full current carrying capacity of ReBCO (Re = rare earths, such as Y, Gd, Eu) materials.

Significant progress in understanding oxygen incorporation processes and identifying influencing factors has been made in the field of solid oxide cells, where the oxygen reduction reaction (ORR) is at the heart of operation.[22–25] The ORR, describing the incorporation of an oxygen molecule into crystal sites, $O_O^\times$, can be simply written as:*

$$O_{2,\text{gas}} + 2v_O^{\bullet\bullet} \rightleftharpoons 2O_O^\times + 4h^\bullet \qquad \text{Eq. 1}$$

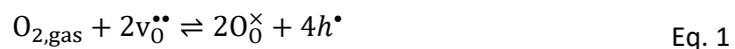

---

* Charges are indicated following the Kröger-Vink notation, with × and • corresponding to neutral and positive charge, respectively, relative to the perfect, defect-free crystal.

with the oxygen vacancy, $v_O^{\bullet\bullet}$, and positively charge electron holes, $h^{\bullet}$. Nevertheless, it consists of a succession of elementary reaction steps, including adsorption, ionization, dissociation, recombination, surface and bulk diffusion of different charged oxygen species as well as electronic charges, as schematically illustrated in Figure 1(a). Consequently, such interfacial, combined electronic and ionic charge transfer reactions are highly complex. Challenging, but progressively systematic studies deploying novel approaches and tools,[22,26,27] strive to illuminate the reaction pathways and identify the slowest elementary mechanism in the chain of reaction steps, thus the rate determining step (RDS), with the goal to improve the performance of catalysts, as well as of oxide electrodes and provide better control in tuning functional properties of mixed ionic electronic conductors (MIECs).

Based on sample dimensions and materials properties, one can distinguish two limiting cases for oxygen incorporation: bulk diffusion and surface exchange limitation.[28] For thin films of thickness $d_{\text{film}}$, oxygen incorporation can be reduced to a 1D problem (along the out-of-plane direction) and one can introduce the dimensionless parameter $L$, also referred to as Biot number:

$$L = \frac{d_{\text{film}} k_{\text{chem}}}{D_{\text{chem}}} \qquad \text{Eq. 2}$$

with the surface exchange and bulk diffusion coefficients, $k_{\text{chem}}$ and $D_{\text{chem}}$, respectively. For $L \ll 1$, any oxygen concentration gradient along the thickness vanishes and the materials response to changes in the oxygen chemical potential (*e.g.* due to changes in the surrounding oxygen partial pressure) is determined by the surface only (surface exchange limited regime). Conversely, for $L \gg 1$, the top surface layer rapidly equilibrates to a new $pO_2$ in the atmosphere, while a concentration profile develops as function of distance to the surface (diffusion limited regime).

As for the YBCO community, which started historically with the analysis of bulk materials, the spotlight was focused for a long time on oxygen diffusion,[29–33] with little consideration of the importance of surface reactions.[34] While this remains plausible for bulk with large diffusion lengths or low density, highly porous materials, the emergence of highly dense, low porosity thin films, as required for superconducting applications, calls for a revision of this picture. In particular, we suggested recently,[35] that oxygen incorporation in YBCO thin films may be exclusively governed by surface reactions, such as ionisation, dissociation and/or recombination of an oxygen surface ion with a bulk vacancy. This is in agreement with a large number of similar thin film perovskite-related oxide materials, where the RDS was identified to rest within the surface,[36–41] and supported by previously reported fast grain-boundary assisted diffusion in YBCO.[42] Generalizing these results however, is problematic, as it may be very much dependent on the specific YBCO specimen. Namely, the materials chemistry itself can strongly depend on the sample type (bulk, powder, thin film), structure (single crystal, polycrystalline, epitaxial) and derived properties, such as density, porosity, surface termination and surface chemistry, number and type of bulk and surface defects and may strongly be affected by the synthesis technique. This results in a wide spread of reported surface exchange coefficients for different YBCO materials, spanning several orders of magnitude, as summarized in Figure 1(b). Other effects however, including surface contamination, catalytic surface activation and detrimental deactivation, have been addressed insufficiently for ReBCO materials, leading to a materials history component which needs to be considered[43–45] and will be subject of this work.

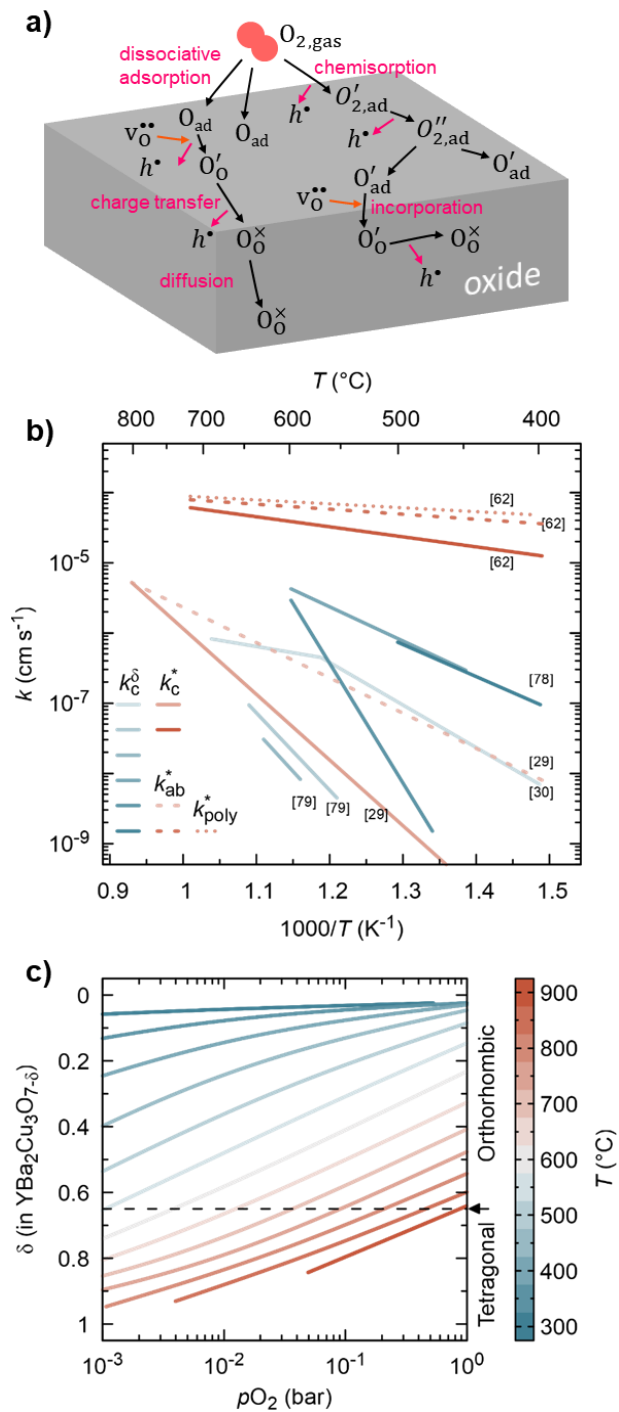

Figure 1: (a) Sketch showing different reaction pathways for the process of oxygen incorporation into an oxide, including various elementary reaction steps, such as adsorption, ionization, dissociation, incorporation and diffusion, involving different defect species such as adsorbed (ad) and incorporated oxygen ions ($O_{ad}, O'_O, O^\times_O$, etc), oxygen vacancies ($v^{\bullet\bullet}_O$) and electron holes ($h^\bullet$), given in Kröger-Vink notation. Notably, a large number of other reaction pathways are similarly possible. (b) Comparison of reported $k$ values obtained via tracer experiments ($k^*$, red) and chemical experiments ($k^\delta = k_{\mathrm{chem}}$, blue) for different crystal directions (solid/dashed lines for surface terminations normal/parallel to the c-axis, dotted line for polycrystalline sample). (c) Oxygen off-stoichiometry, $\delta$, of bulk YBa$_2$Cu$_3$O$_{7-\delta}$ as function of $pO_2$ at various temperatures, adapted from literature.[46]

While the above discussed phenomena are of kinetic nature, the equilibrium oxygen stoichiometry in an oxide is defined thermodynamically via its oxygen chemical potential and is as such a function of oxygen partial pressure as well as temperature, with inverse effects on the oxygen concentration.[47] That means increasing the $pO_2$ leads to a higher oxygen stoichiometry, while higher temperatures increase the configurational entropy and thus result in a lower overall oxygen concentration, due to a higher number of point defects, such as oxygen vacancies. An oxygen non-stoichiometry diagram for bulk YBCO adapted from literature,[46] is shown in Figure 1(c). As superconducting properties improve with increasing oxygen content, low annealing temperatures are favourable to obtain $\delta \approx 0$. However, the kinetic process of oxygen incorporation is thermally activated, resulting in a trade-off conflict between fast kinetics at high $T$ and a small off-stoichiometry a low $T$. It may therefore seem beneficial to start the oxygenation process at high temperatures, where oxygen vacancies are rapidly filled and subsequently decrease the temperature to slowly occupy the remaining vacant oxygen sites. In fact, a similar process may be followed by some commercial coated conductor tape manufacturers.[33,48] However, this intuitive approach omits possible degradation processes, such as surface poisoning, reconstruction, formation of extended bulk defects, etc., which are commonly observed in oxide electrodes for high temperature applications (SOFC, SOEC),[49–53] as well as HTS materials[54] and have to be considered in designing an optimized oxygenation process.

Here, we use electrical conductivity relaxation (ECR) measurements to study the oxygen activity of bare and Ag decorated YBCO thin films as function of temperature and history, completed by a microstructural analysis. We investigate for kinetic degradation and catalytic activation phenomena and develop a possible, yet simplified reaction pathway scenario of oxygen kinetics in YBCO, with the goal to provide guidelines for improved oxygenation treatments.

## 1. Experimental

Thin films studied within this work were synthesized using the chemical solution deposition (CSD) technique,[55] whereas anhydrous TFA precursor salts were dissolved in trifluoroacetic acid, trifluoroacetic anhydride and acetone to obtain a 0.25 mol (with respect to Y) YBCO-TFA solution. The solution was deposited by spin-coating on 5×5 mm² $LaAlO_3$ (00l) (LAO) and $StTiO_3$ (00l) (STO) single crystal substrates (Crystec). The subsequent pyrolysis and growth processes are described in detail elsewhere.[56,57] The highly epitaxial growth was confirmed and the *c*-axis lattice parameter was determined using a high resolution Discover D8 Bruker diffractometer (X-ray energy= 8.049 keV). Electrical conductivity was measured in Van der Pauw configuration using small excitation currents (± 100 μA). The low temperature resistivity, $\rho$, was measured in a Physical Property Measurement System (PPMS, Quantum Design). Electrical measurements were averaged over two permutations of the electrical contacts and positive and negative excitation currents in DC mode (100 μA). The critical current density was obtained via SQUID magnetometry (Quantum Design), as explained elsewhere.[16]

For *in situ* electrical measurement, samples were mounted onto a ceramic sample holder using thin Ag wire glued with high temperature Ag paste onto the top corners and placed in the centre of a calibrated 22 mm tube furnace, equipped with a calibrated thermocouple in close proximity to the sample. For better electrical contacts, squared 700×700 μm and 100 nm thick Ag electrodes were fabricated in the four top corners of the samples using optical lithography (DURHAM) and argon DC sputtering (TSST) at ICMAB's cleanroom facilities. The same process was used to prepare Ag decorated YBCO thin films (YBCO|Ag). To enable electrical measurements, this 100 nm Ag coating was micro-patterned, as shown in Figure 4(c).

The oxygen exchange activity was studied using electrical conductivity relaxation (ECR) measurements. Therefore, the response of the electrical conductivity, $\sigma$, of a thin film to a rapid change in the atmospheric $pO_2$ is recorded over time and the transient from the initial to the final equilibrium state is analysed to obtain kinetic insight. The modulation of the materials conductivity is caused by changes in the oxygen stoichiometry, $c$, and thus the charge carrier concentration via the oxygen reduction reaction (Eq. 1). Reduction and oxidation processes are triggered by jumps to lower and higher $pO_2$, respectively. Oxygen partial pressures in the range of 1 mbar and 1 bar were established by mixing high purity gases of $O_2$ and $N_2$ using gas flow controllers. All experiments were carried out at atmospheric pressure. Rapid gas atmosphere changes were enabled using high gas flows (0.6 l min$^{-1}$) and automatic valves controlling the switching process via a digital relay. Heating and cooling ramps were set between 3 and 10 °C min$^{-1}$ and were always executed in oxygen rich atmosphere. After reaching the desired temperature and stabilization for several minutes, the reduction-oxidation cycle was initiated.

For thin films, the oxygen diffusion problem can be reduced to 1D, with a linear surface exchange term as boundary condition at the top surface only. In the case that the overall reaction is limited by a surface reaction (*i.e.* $D_\text{chem} \gg d_\text{film} k_\text{chem}$), the solution of Fick's diffusion law for a plane sheet sample geometry can be approximated as:

$$\frac{c(t) - c_\infty}{c_0 - c_\infty} \cong \frac{\sigma(t) - \sigma_\infty}{\sigma_0 - \sigma_\infty} \cong e^{-\frac{t}{\tau}} \qquad \text{Eq. 3}$$

with the relaxation time, $\tau = V/(A k_\text{chem})$, which relates the characteristic time of the reaction process with the surface exchange coefficient, $k_\text{chem}$, the thin film volume, $V$, and exposed surface area, $A$. For the applied model of dense, thin films of rectangular cuboid shape, the ratio $V/A$ reduces to the film thickness, $d_\text{film}$, and consequently $\tau = d_\text{film}/k_\text{chem}$. The sub-indices 0 and ∞ indicate the initial and saturation values of the oxygen concentration and the conductivity, respectively. The validity of Eq. 3 is limited for small changes in the oxygen stoichiometry and thus *small* jumps in $pO_2$, for which a linear correlation with the conductivity can be assumed, *i.e.* $\sigma \propto c$ and a simple, linear kinetic regime prevails.[58]

## 2. Results

### 2.1. Onset of oxygen incorporation

At elevated temperatures, the YBa$_2$Cu$_3$O$_{7-\delta}$ phase is thermodynamically stable only within a narrow ($T$, $pO_2$) window. Grown under these conditions, YBCO is highly oxygen deficient, *i.e.* $\delta \gg 0$. Thus, independent of the growth technique, oxygen post-growth treatments are required to reduce the oxygen off-stoichiometry, thereby introducing charge carriers for the superconducting phase to emerge and flourish. To study the initial oxygen uptake of YBCO thin films, the freshly synthesized YBCO material was cooled from the deposition temperature in dry growth atmosphere ($pO_2 = 0.2$ mbar) without an increase of the $pO_2$ during the cooling. This results in an oxygen off-stoichiometry of about $\delta \approx 0.5 - 0.6$, as estimated from the superconducting transition below 40 K (*cf.* Figure 2(d)).[14] In this strongly reduced state, electrical contacts were fabricated on the top corners for *in situ* resistivity and ECR measurements. Additional batches of samples were prepared with a thin silver decoration layer on the top surface (YBCO|Ag, see Figure 4(c)).

The subsequent annealing in 1 atm of oxygen is characterised by a strong chemical driving force for oxygen incorporation. The $\rho(T)$ curves of a bespoken initial heating in oxygen are shown in Figure 2(a) for an uncoated and an Ag decorated YBCO sample. The samples exhibit approximately metallic like behaviour with a linear increase of $\rho$ up to a certain temperature, $T_\text{onset}$, followed by a strong decrease at higher $T$. The peak is shifted by approximately 50 °C to lower temperatures for the YBCO|Ag system, as compared to the bare YBCO surface. This effect was observed systematically for the initial oxygenation in all studied samples, with a clear shift of $T_\text{onset}$ to lower temperatures upon Ag coating, as summarized in Figure 2(b). Furthermore, the temperature of the maximum in $\rho$ depends on the deployed heating rate, as shown in Figure 2(c), pointing towards the involvement of kinetic processes.

Similar $\rho(T)$ curves were reported previously for YBCO,[59,60] where it was proposed that this effect is linked to an oxygen ordering mechanism inside the bulk, but the exact origin remained unclear. Based on our findings, we suggest that the maximum in $\rho(T)$ is related to the onset of oxygen incorporation processes and as such a highly relevant indicator for the minimum temperature for an optimised oxygen treatment. The following strong decrease in $\rho$ above approx. 250 – 320 °C is caused by the increase of the free charge carrier density via the oxygen reduction reaction, which is catalytically enhanced through the Ag decoration, as discussed in more detail in the next section. To validate this assumption, we oxygenated bare and Ag coated YBCO for 5 h at 300 °C and analysed the resulting low temperature physical properties. We selected 300 °C, as it approximately lies in between the $T_{onset}$ for bare and Ag coated YBCO.

The low temperature resistivity and critical current density curves are shown in Figure 2(d & e), respectively. The YBCO|Ag specimen exhibits high oxygen doping, characterised by lower resistivity values with high linearity in temperature,[61] a high critical temperature (drop to zero resistance at around 89 K) and adequate $J_c$ values for CSD-TFA grown YBCO samples.[21,55] On the other hand, the oxygen content in bare YBCO was only slightly enhanced compared to the as grown state, indicated by an increase of the critical temperature from below 40 K to about 64 K and a slightly reduced $\rho(T)$. The low $J_c$ values of the poorly oxygenated bare YBCO sample underline the importance of oxygen doping to achieve excellent superconducting properties. This data is in agreement with our assumption, that the drop in $\rho$ above $T_{onset}$ is linked to the incorporation of oxygen and confirms that silver decoration enables oxygen treatments at reduced temperatures compared to bare YBCO, where kinetic processes and oxygen incorporation are activated at slightly higher temperatures ($T_{onset} \approx 320$ °C).

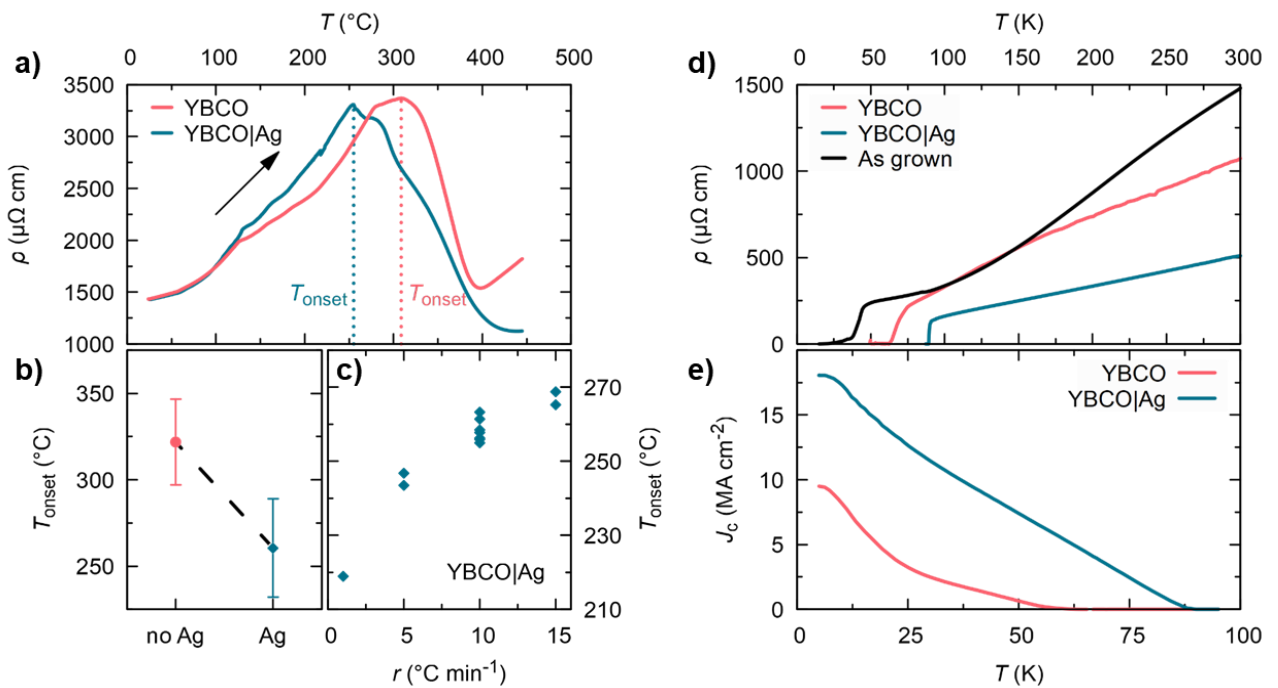

Figure 2: Investigation of the onset of oxygen incorporation in YBCO thin films using in situ resistivity measurements: (a) Temperature dependence of $\rho$ for uncoated and Ag decorated YBCO samples during the first heating in oxygen rich atmosphere (1 atm). The maximum in $\rho$ marks the onset temperature for oxygen incorporation, $T_{onset}$. (b) Onset temperature of oxygen incorporation for bare and Ag decorated YBCO averaged over several samples (at a heating ramp of 10 °C min$^{-1}$). (c) Dependence of $T_{onset}$ on the heating rate for silver coated samples. (d) Low temperature resistivity and (e) critical current density of bare and Ag coated YBCO oxygenated at 300 °C. Panel (d) additionally shows an as grown YBCO sample without oxygen treatment. Note the absolute temperature scale in (d & e).

## 2.2. Oxygen kinetics of YBCO thin films

In the previous section, we found an unexpected low onset temperature of oxygen incorporation. To determine the oxygen exchange activity of YBCO thin films in a systematic manner, we deployed the ECR technique following two different, time-mirrored protocols, by either starting at high and going to low temperatures (H2L), or in inverse direction going from low to high temperatures (L2H), spanning a temperature range from 375 to 600 °C, with reduction-oxidation cycles every 25 °C. This analysis was performed for bare, as well as silver decorated YBCO thin films.

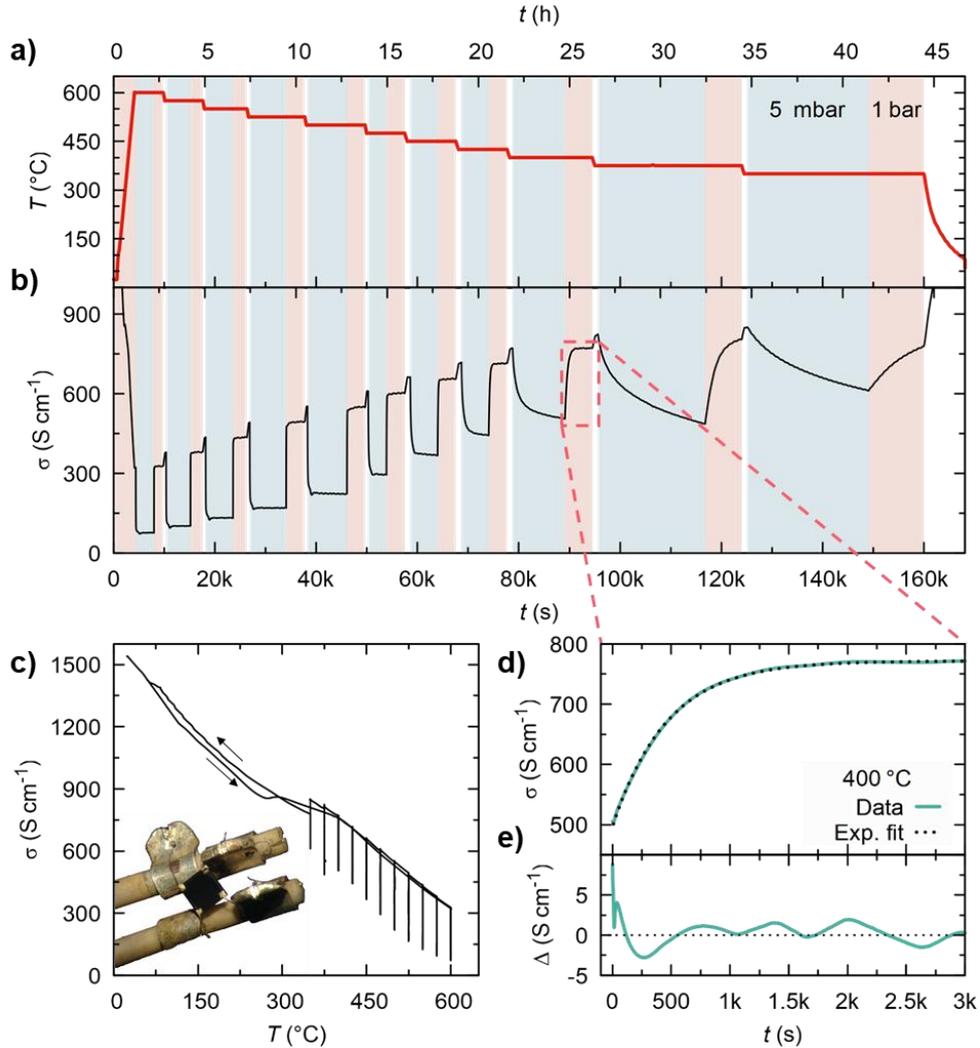

Figure 3: Electrical conductivity relaxation measurements from high to low (H2L) temperatures for a YBCO|Ag specimen: (a) Temperature and (b) conductivity as function of time. Background coloring indicates the oxygen partial pressure (blue: 5 mbar, red: 1 bar, white: 1 bar during heating stages). (c) Conductivity as function of temperature. Note that this sample was cooled in oxygen rich atmosphere after growth and therefore does not exhibit a strong feature around 300 °C, corresponding to the onset of oxygen incorporation. Inset shows optical image of the sample holder. (d) Oxidation step at 400 °C (marked with a rectangular in (b)), fitted using an exponential curve, with the corresponding data-fit deviation in (e).

The temperature and $pO_2$ profile as well as the evolution of the conductivity for a H2L experiment are presented in Figure 3(a-b). The corresponding $\sigma(T)$ curve is depicted in Figure 3(c), with the typical *spikes* for each cycle. Notably, the difference in conductivity between the initial heating and the subsequent cooling steps is small. A kinetic analysis is performed for the isothermal reduction-oxidation steps to obtain surface exchange coefficients. The oxidation transients can be well fitted using to a single exponential decay, as given in Eq. 3. An example is shown in Figure 3(d & e), showing the very high fit quality. Generally, reduction processes could not be adequately fit using a single exponential term. The introduction of a second, parallel exponential process, as commonly done in literature,[38,62–64] allows to appropriately model experimental data. However, for clarity of this manuscript, we restrict the discussion to oxidation processes, sufficiently resembled using a single exponential process.

The obtained surface exchange coefficients for oxidation are shown in Figure 4(a & b) for the H2L and L2H process, respectively, for bare and Ag coated YBCO. We find thermally activated behaviour for all four samples, following approximately an Arrhenius-law with a single activation energy, $E_A$. Throughout the analysed temperature range and for both measurement directions, YBCO|Ag exhibits faster kinetics, as compared to bare YBCO. The fact that the simple decoration of the YBCO surface with Ag modifies the oxygen incorporation rate, strongly suggests that the ORR for the studied bare YBCO thin films is limited by surface reactions, as reported previously in our work.[65] Silver can be expected to facilitate the ionization of $O_2$ by providing easy electron charge transfer to the oxygen molecule and thereby lowering its dissociation energy.[66–68] This job sharing mechanism is sketched in Figure 4(d) and makes the reaction kinetically more favourable, leading to faster exchange coefficients.

The dash red lines in Figure 4(a & b) mark the approximate boundary for the oxygen incorporation being limited by a surface process. It is based on reported values for YBCO diffusion coefficients,[69] via $k_{\text{limit}} = D_{\text{chem}}/d_{\text{film}}$. For higher $k_{\text{chem}}$ values, bulk diffusion would start co-limiting the exchange process and finally become the governing factor. We note, that also for the silver coated 250 nm thick YBCO films the values fall below $k_{\text{limit}}$. Additionally, oxidation curves of YBCO|Ag can be well modelled using a single saturation time over the full course of the relaxation process, *cf.* Figure 3(d & e), including small times after the change in atmosphere close to $t \approx 0$, which would not be expected for a diffusion (co-)limited process.[58,70] We therefore expect, that all studied samples fall into the surface limited regime, including the ones with catalytic Ag decoration.

Comparing the H2L and L2H processes, one finds a strong dependence of bare YBCO on the direction of the temperature profile, in both magnitude of the observed $k_{\text{chem}}(T)$ and (apparent) activation energies (switching between 0.7 and 3.4 eV). For YBCO|Ag we obtain very similar results with reasonably matching $k_{\text{chem}}$ values and activation energies of about 1 eV in both measurement directions. Interestingly, for pristine YBCO the H2L process leads to faster kinetics at higher temperatures, while the L2H approach preserves higher activity at low $T$. Correspondingly, the gap in $k_{\text{chem}}$ values between silver coated and pristine YBCO is closer at the beginning of the measurements (independent of a high or low starting temperature) and increases with annealing time. This suggests a deactivation of exchange kinetics over time, commonly found for similar perovskite oxides used as electrode materials in solid oxide cell applications.[49,50,71]

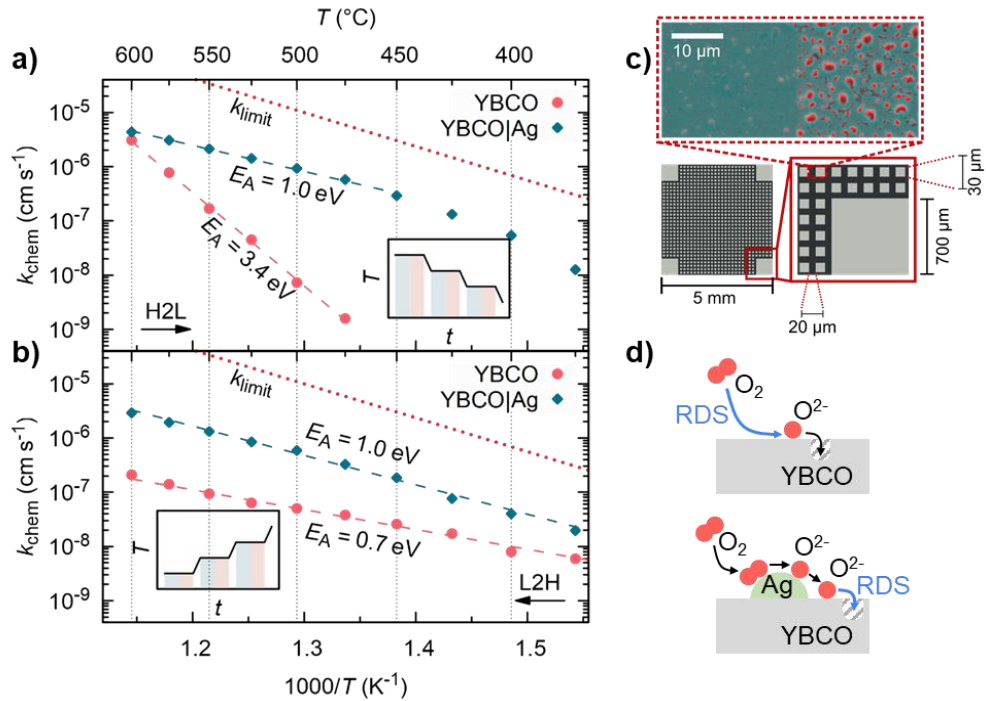

Figure 4: Arrhenius plots of surface exchange coefficient, $k_{chem}$, for oxidation steps from 5 to 1000 mbar of bare and Ag decorated CSD-YBCO thin films obtained via ECR measurements starting at (a) high temperatures (H2L) and (b) low temperatures (L2H). The insets schematically show the temperature profile and $pO_2$ steps (blue shaded areas correspond to low, red areas to high $pO_2$). The red dotted lines mark the approximate upper limit for $k_{chem}$ for a surface limited process (see main text). (c) Schematic of the employed micro-patterned Ag coating layer and false color SEM image (combined secondary and back scattered electron mode) showing dewetting of Ag pad into µm-sized islands (above 300 °C). (d) Expected role of the catalytic activity of silver by providing a job sharing mechanism for oxygen incorporation.

We studied these deactivation processes qualitatively by performing consecutive isothermal reduction (blue) and oxidation (red) steps at 450 °C, while monitoring the evolution of the conductivity, as shown in Figure 5(a & b) for a fresh set of bare and Ag decorated YBCO samples. Starting from fast relaxation curves, the saturation times for reaching equilibrium drastically increase for uncoated YBCO with increasing total annealing time (indicated by going from light to dark colours), Figure 5(a). The temporal evolution of the surface exchange coefficient for oxidation is depicted in Figure 5(c), revealing a strong decrease of almost two orders of magnitude within 17 h of annealing. A small deactivation is also observed for YBCO|Ag, however, at a much lower rate. Silver therefore not only catalytically enhances the oxygen exchange activity of YBCO (compare initial $k_{chem}$ values at 3k s), but also prevents kinetic degradation.

On the other hand, the saturation conductivity values, $\sigma_\infty$, obtained by raw data fitting, remain approximately constant throughout the measurement for both sample types as well as for oxidation and reduction processes, as shown in Figure 5(d). This means that the electronic structure is not deteriorated and the thermodynamic saturation value is not modified, opposite to the kinetic oxygen incorporation processes.

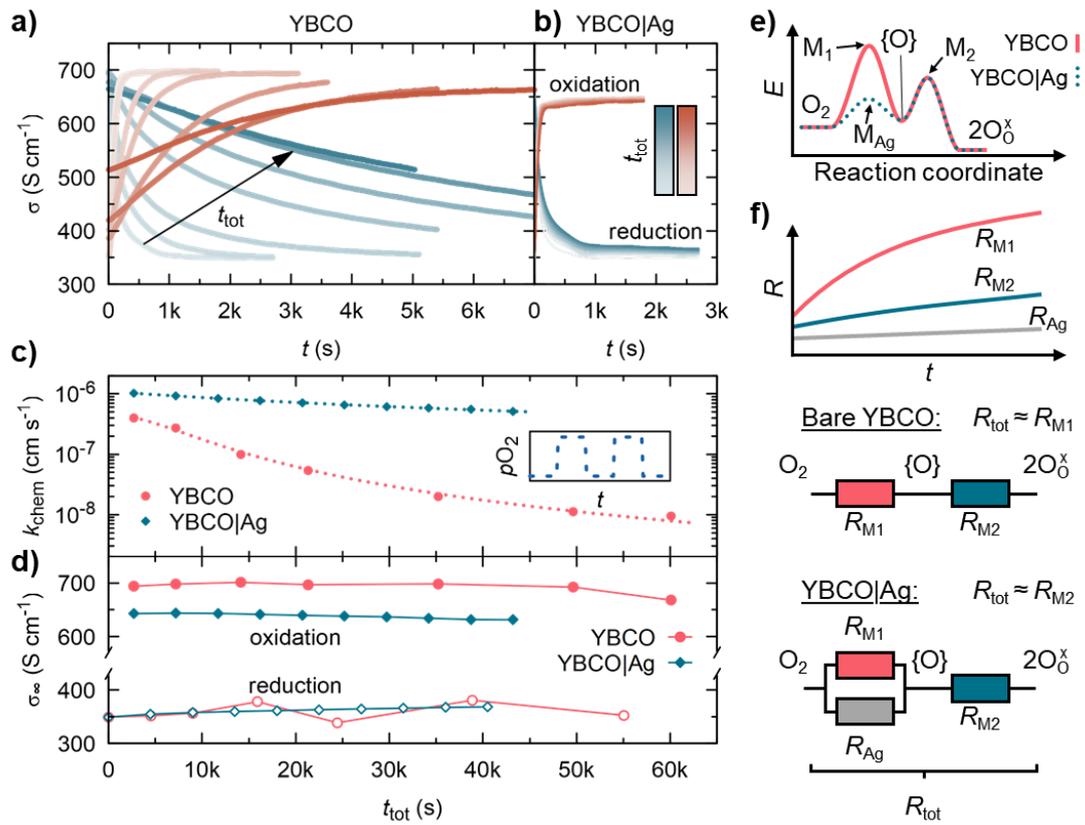

Figure 5: Consecutive reduction (blue, from bright to dark) and oxidation (red) steps at 450 °C followed using electrical conductivity for (a) uncoated and (b) Ag coated CSD-YBCO. (c) Evolution of the oxidation exchange coefficient, $k_{chem}$, with the total annealing time at 450 °C, revealing a strong deactivation of the surface activity in bare YBCO. (d) Evolution of the saturation conductivity for oxidation and reduction, obtained by fitting (corresponding to $t = \infty$). (f) Electrical analogue of the ORR based on the representation of the elementary mechanisms $R_{M1}$, $R_{M2}$ and $R_{Ag}$ using series and parallel resistances, and a schematic of their evolution with time.

Based on our observations, we can propose a possible, simplified ORR reaction mechanism, as shown in the reaction energy diagram in Figure 5(e). In bare YBCO, oxygen incorporation may be limited by either adsorption, ionisation and/or dissociation of $O_2$, which we can combine in the reaction mechanism, $M_1$, and leads to the (unknown) oxygen intermediate, {O}. This transition state evolves further to reach the final oxygen bulk level of $O^{2-}$, possibly via additional ionization steps and the recombination with a bulk vacancy and oxygen bulk diffusion, summarized as $M_2$. As schematically shown in Figure 5(e), the activation energy of $M_1$ is higher than $M_2$, therefore limiting the overall reaction rate.

As silver coating can be thought to catalytically activate oxygen adsorption, ionisation and/or $O_2$ dissociation, it provides a faster, alternative pathway, $M_{Ag}$, running in parallel to $M_1$ with a strongly reduced activation energy. In this case, $M_2$ becomes the rate determining step for the ORR. Given the small film thickness, small $k_{chem}$ values (compared to $k_{limit}$ as shown in Figure 4), the observed single exponential decay for oxidation steps, and a preliminary analysis of the reaction order of reaction rates[72] (a more thorough analysis is currently in preparation), we anticipate that also for Ag coated YBCO a surface reaction is limiting the overall oxygen incorporation rate. RDS candidates for $M_2$ include surface diffusion processes of charged oxygen species, (additional) ionization steps or the recombination of an oxygen ion with a bulk oxygen vacancy.

To persuade the reader of the proposed reaction pathway with respect to the observed ageing behaviour, we can qualitatively describe the situation using an electrical resistance analogue, where the different mechanisms are represented by series and parallel resistances, as shown in Figure 5(f). Bigger resistances correspond to higher activation energy barriers (and thus slower kinetics). As for bare YBCO $R_{M1} \gg R_{M2}$, the total resistance of this reaction is determined by $R_{M1}$, and the observed deactivation rate is linked to an increase of resistance of $R_{M1}$ with time. In the case of Ag coated YBCO, $R_{M1}$ is bypassed via the small parallel resistance provided via the catalytic pathway of silver, $R_{Ag}$. The total resistance is thus governed by $R_{M2}$ and our kinetic measurements become sensitive to the degradation rate of $R_{M2}$. As the observed deactivation rates strongly differ, we speculate that they are rooted in different chemical and/or structural processes. However, it is highly challenging to identify the specific cause for these degradation mechanisms, as manifold processes down to the atomic level could play a role, including a restructuring of the surface, chemical ageing, the build-up of surface contaminations, *etc.*, and a detailed analysis goes beyond the scope of this work.

It is noteworthy, that qualitatively similar degradation phenomena were observed in 200 nm thick YBCO thin films grown by pulsed layer deposition,[72] suggesting that ageing phenomena are a more general problem for YBCO surfaces, independent of the synthesis routine. This calls for further detailed studies to provide a clear picture of the origin of these processes.

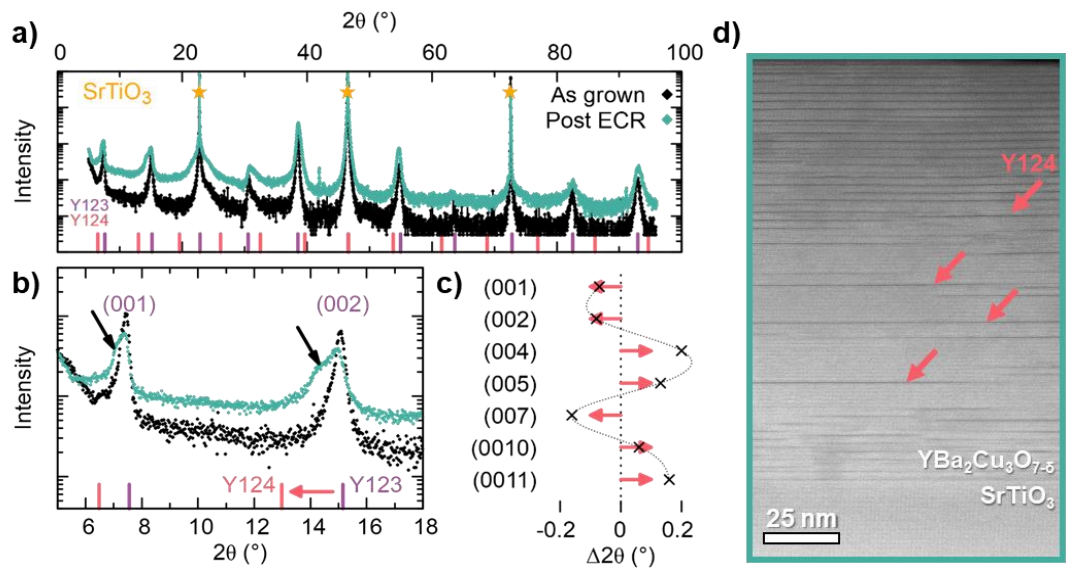

Figure 6: XRD analysis of as deposited sample and specimen after extended thermal cycling (post ECR). Full range (a) and magnified region (b) of the x-ray diffraction spectra. (00l) reference phases for YBa$_2$Cu$_3$O$_{7-\delta}$ (Y123)[73] and YBa$_2$Cu$_4$O$_{8-\delta}$ (Y124)[74] are marked at the bottom. Black arrows in (b) indicate the appearance of a shoulder. (c) Y123 (00l) peak shifts in $2\theta$ after annealing compared to as deposited state. The oscillating curve is a guide to the eye. Pink arrows in (b & c) mark the direction of the closest Y124 peak. (d) TEM image of post ECR YBCO sample. Arrows indicate the presence of Y124 intergrowths.

### 2.3. Microstructural effects of oxygen treatments

We concluded our experimental work with the bulk analysis using x-ray diffraction of YBCO thin films at two different stages: as grown and after extended ECR analysis (post ECR). The oxygen content strongly influences the structural parameters, including a tetragonal to orthorhombic phase transition and the shrinkage of the c-axis lattice parameter with decreasing oxygen off-stoichiometry. To minimize associated effects, the as grown sample was exposed to an oxygen rich atmosphere during cooling (in contrast to the samples discussed in the previous sections), whereas the oxygen pressure inside the furnace was increased to 1 bar below 600 °C. In Figure 6(a) we compare the same sample with similar overall oxygen content, but different thermal history, *i.e.* just after the deposition and after extended ECR analysis (post ECR),which was performed between 350 and 600 °C with a total annealing time of about 45 h. First, the data confirms the epitaxial, *c*-axis oriented growth of YBCO on the SrTiO$_3$ single crystal substrate, as we only observe the (00l) peak family of YBa$_2$Cu$_3$O$_{7-\delta}$ (Y123) in the diffraction pattern. No impurity phases are detected. Second, the annealing does neither lead to the rise of isolated peaks, nor to broad shifts of individual reflections to different angles. However, subtle changes are observed via the formation of shoulders and small shifts in 2Θ of the Y123 peak family, as magnified in Figure 6(b) (marked with black arrows). A commonly found extended structural defect in YBCO is the formation of CuO intergrowths, corresponding to the YBa$_2$Cu$_4$O$_{8-\delta}$ (Y124) phase, so called stacking faults (SF). The 123 stoichiometry is maintained, as the SFs are highly defective including numerous Cu and O vacancies.[75] Due to the second CuO-chain, Y124 has a larger *c*-axis cell parameter and correspondingly its (00l) peaks are shifted to lower angles, *cf.* pink bottom markers in Figure 6(a & b). In agreement with previous reports,[76–79] we do not observe the occurrence of coherent reflections of the Y124 phase, but we systematically detected a small 2Θ shift of the Y123 peaks towards the closest Y124 reflection, as shown in Figure 6(c). Note that the pink arrows mark the direction of the closest Y124 peak, not necessarily the one with the same miller indices.

This oscillating positive and negative shift hints towards the formation of stacking faults, as confirmed by TEM cross section characterisation, given in Figure 6(d). We find a high density of additional and extended Y124 intergrowths, easily identified due to their high contrast (marked by arrows). These defects were observed independent of Ag coating and thus are not expected to influence the deactivation of the surface kinetics. Nevertheless, SFs influence the pinning behaviour in the superconducting state, which can be either beneficial or detrimental, depending on their density and length, the operation temperature and external magnetic field conditions (strength and orientation).[78] Therefore, their formation during (long) thermal annealing processes (in oxygen rich environment) has to be taken into account for the optimization of superconducting properties.

## 3. Conclusions

We analyzed the oxygen surface kinetics of YBCO thin films grown by chemical solution deposition using electrical conductivity relaxation measurements. Oxygen incorporation in pristine YBCO was found to be activated already at low temperatures (≈ 320 °C), however, suffers from rapid degradation. Surface decoration using Ag micro-islands catalyzes oxygen surface reactions, shifts its onset to lower temperatures and protects from deactivation via an alternative oxygen incorporation pathway based on a job-sharing mechanism. Our findings suggest that oxygen incorporation in all studied thin films is limited by surface reactions and we propose a possible, simplified reaction pathway scenario. Additionally, we observed the formation of extended bulk defects in the form of long CuO stacking faults after extended thermal cycling. It is therefore important to establish a well-designed oxygenation process in terms of temperature and time to allow for good superconducting properties at reduced processing costs and avoid materials degradation processes, highlighting its relevance for both, the academic and industrial superconductivity community.

## Acknowledgments

Authors acknowledge funding from EU COST actions OPERA (CA20116) and SUPERQUMAP (CA-21144), the Spanish Ministry of Science and Innovation and the European Regional Development Fund, MCIU/AEI/FEDER for SUPERENERTECH (PID2021-127297OB-C21), "Severo Ochoa" Programs for Centers of Excellence in R&D Matrans42 CEX2023-001263-S. They also thank the Catalan Government for 2021 SGR 00440. Authors also thank the Scientific Services at ICMAB and ICN2 Electron Microscopy Division.

## Data Availability Statement

The data that support the findings of this study will be made available in zenodo at https://doi.org/10.5281/zenodo.15333534/

## Declaration of Competing Interest

## Author Contributions

A.S. developed the original idea of this manuscript and designed the methodology and experimental studies with contributions from T.P. A.S. performed the experimental work, analyzed the data and prepared the manuscript. T.P. and A.B. acquired the funding for research and personnel. All authors contributed to the revision of the manuscript.